\documentclass[pdflatex,sn-mathphys]{sn-jnl}% Math and Physical Sciences Reference Style
\jyear{2022}%

\theoremstyle{thmstyleone}%
\theoremstyle{thmstyletwo}%

\theoremstyle{thmstylethree}%

\usepackage{xcolor,colortbl}

\begin{document}

\title[An approach to hummed-tune and song sequences matching]{An approach to hummed-tune and song sequences matching}

%%=============================================================%%
%% Prefix	-> \pfx{Dr}
%% GivenName	-> \fnm{Joergen W.}
%% Particle	-> \spfx{van der} -> surname prefix
%% FamilyName	-> \sur{Ploeg}
%% Suffix	-> \sfx{IV}
%% NatureName	-> \tanm{Poet Laureate} -> Title after name
%% Degrees	-> \dgr{MSc, PhD}
%% \author*[1,2]{\pfx{Dr} \fnm{Joergen W.} \spfx{van der} \sur{Ploeg} \sfx{IV} \tanm{Poet Laureate} 
%%                 \dgr{MSc, PhD}}\email{iauthor@gmail.com}
%%=============================================================%%

\author*[1]{\fnm{Bao Loc}\sur{Pham}}\email{phambaoloc163@gmail.com}

\author*[1]{\fnm{Huong Hoang} \sur{Luong}}\email{huonghoangluong@gmail.com}
\equalcont{These authors contributed equally to this work.}

\author[1]{\fnm{Thien Phu} \sur{Tran}}\email{hostcode0301@gmail.com}
\equalcont{These authors contributed equally to this work.}

\author[1]{\fnm{Hoang Phuc} \sur{Ngo}}\email{phucnnh21@gmail.com}
\equalcont{These authors contributed equally to this work.}

\author[1]{\fnm{Hoang Vi} \sur{Nguyen}}\email{hoangvi0206@gmail.com}
\equalcont{These authors contributed equally to this work.}

\author[1]{\fnm{Thinh} \sur{Nguyen}}\email{thinhnpwork111@gmail.com}
\equalcont{These authors contributed equally to this work.}

\affil*[1]{\orgdiv{IT Department}, \orgname{FPT University}, \orgaddress{\street{Nguyen Van Cu}, \city{Can Tho}, \postcode{90000}, \state{Can Tho}, \country{Vietnam}}}

%%==================================%%
%% sample for unstructured abstract %%
%%==================================%%

\abstract{% Melody stuck in your head, also known as "earworm", is tough to get rid of it unless you listen to the song or sing it, but what if you can not find the name. The task of recognizing a song name base on humming is not easing for a human being. Adapt from Hum2Song Zalo AI Challenge 2021, the competition is about querying the name of a song by giving user’s humming sound, similar to Google’s hum to search. The data come from the challenge contains 1.3GB of trainning data, 539MB of public test, and 12.5GB of private test (which use for the competition leaderboard). The methodology adapts from the first-place solution, this paper contains go in detail about preprocessing the audio which origin form of mp3 into usable form for training and inference, creating a model which contains state of the art deep learning backbone for feature extraction from song and hum, with using Faiss module for searching matching.

Melody stuck in your head, also known as "earworm", is tough to get rid of, unless you listen to it again or sing it out loud. But what if you can not find the name of that song? It must be an intolerable feeling. Recognizing a song name base on humming sound is not an easy task for a human being and should be done by machines. However, there is no research paper published about hum tune recognition. Adapting from Hum2Song Zalo AI Challenge 2021 - a competition about querying the name of a song by user's giving humming tune, which is similar to Google's Hum to Search. This paper covers details about the pre-processed data from the original type (mp3) to usable form for training and inference. In training an embedding model for the feature extraction phase, we ran experiments with some states of the art, such as ResNet, VGG, AlexNet, MobileNetV2. And for the inference phase, we use the Faiss module to effectively search for a song that matched the sequence of humming sound. The result comes at nearly 94\% in MRR@10 metric on the public test set, along with the top 1 result on the public leaderboard.
}

%%================================%%
%% Sample for structured abstract %%
%%================================%%


%%\pacs[JEL Classification]{D8, H51}

%%\pacs[MSC Classification]{35A01, 65L10, 65L12, 65L20, 65L70}

\maketitle

\section{Introduction}\label{introduction}
% \subsection{About the problem}
% Adapt from Zalo AI Hum2Song 2021\cite{ref_url1}, the problem is to find the matching song sequence given the hummed tune. Metric for competition is Mean reciprocal rank (MRR). The real life application you can think of is Shazam\cite{ref_url6} and Hum to search (Google). 

% This paper cover the data preprocess, training and inference part.

% \subsection{About the Data}
% Data is come from the Zalo AI Hum2Song 2021 competition. The training data is 1.3GB of size. Contains of 2901 original song sequences from the studio, 2901 hummed tuned. The public test is 539MB of size, this data contains no song sequences or hummed tune from the training data. The public test data is for showing the public leader board during the time of competition. Final is the private test data 12.5GB, this data is for the final leader board.

% All the data is mp3 format. This paper cover the preprocess part for the training and inferencing.

Recently, with the development of multimedia on the advance of mobile technology, people listen to music more than ever. People spend most of their time listening to music, while shopping, driving, or studying. As music is listened to more often, we have more "earworm" than ever. What if you want to listen to "that song" again, how can you find it when you do not know the song's name? To address that issue with current on-the-market products, we got Shazam\cite{ref_url6} and Google's Hum to search\cite{ref_url4}. 

Shazam, released in 2002, is an application that allows you to search for a song's name by letting it listen to music sequences. Nevertheless, Shazam\cite{ref_url6} has a drawback: it can only search by receiving a song recorded from the studio. With other variants such as remix songs or covers, Shazam does not always guarantee having relative accuracy as the original recorded song, which means when coming to human's humming sound, the program cannot return a correct result.

Google's Hum to search\cite{ref_url4}, released in 2020, has the ability to return some of the matching songs that come from user's hummed tunes. It returns some of the most likely songs to the humming tune. All popular songs from the 80s to now trending music have really good accuracy.  

All those mentioned products do not work with Vietnamese songs because the popularity of music genres on the internet is mostly written in English, Spanish, Korean, or Japanese.
% In addition, Vietnam has its own entertainment market with unique musical tastes.
As a result, Vietnamese people need a search engine for their music, which still can not be fulfilled by Shazam or Google's Hum to search. Fortunately, Zalo - the most popular cross-platform instant messaging application in Vietnam, hosted The Zalo AI Challenge 2021\cite{ref_url1}. In that contest, there was a challenge named Hum2Song, which asked the candidates to develop a Machine Learning model to look for a song using a humming tune. Our purpose for this paper is to present the methodology for solving a music matching problem with the Hum2Song\cite{ref_url1} challenge's data.

There are some researches for Vietnamese data: \cite{ref_article23}, \cite{ref_article24}, \cite{ref_article25}, \cite{ref_article26},etc. All of them focus on the classification, in this paper, we will cover about the Vietnamese music searching pipeline.

This paper consists of 5 sections. The next section is about Related works[\ref{relatedworks}]. The Methodology section[\ref{methodology}] will cover the details of the data preprocessing, training and inference pipeline. The Experiments section[\ref{experiments}] will list all our experiment results. Finally, in the Conclusion section[\ref{conclusion}], we conclude our paper and summarize other ways to improve or further research for better results. 

\section{Related Works}\label{relatedworks}
There already some researches for specific tasks in music. There are music classification: 
\cite{ref_article14}, \cite{ref_article15}, \cite{ref_article16}. And cover songs identification : \cite{ref_article18}, \cite{ref_article19}, \cite{ref_article20}, etc. Most of them are supervised learning, but labeling data is a time-consuming job. Thankfully, there are many other methods, such as "contrastive learning methods to train neural networks"\cite{ref_article21}. "The idea is to make the distance of sequences from the same song close to each other, and sequences from different songs must be far apart."\cite{ref_article21}
% "Automatic tagging using deep convolutional neural networks"\cite{ref_article14}, "Samplelevel deep convolutional neural networks for music auto-tagging using raw waveforms"\cite{ref_article15}, "Pre-Trained Convolutional Neural Networks for Music Audio Tagging"\cite{ref_article16}, "Transfer Learning by Supervised Pre-Training for Audio-Based Music Classification: Semantic Scholar"\cite{ref_article17}. And cover songs identification : "Similarity Learning for Cover Song Identification Using Cross-Similarity Matrices of Multi-Level Deep Sequences"\cite{ref_article18}, "Key-Invariant Convolutional Neural Network toward Efficient Cover Song Identification"\cite{ref_article19}, "Learning a Representation for Cover Song Identification Using Convolutional Neural Network"\cite{ref_article20}, etc. Most of them are supervised learning, but labeling data is a time-consuming job. Thankfully, there are many other methods, such as "contrastive learning methods to train neural networks"\cite{ref_article21}. "The idea is to make the distance of sequences from the same song close to each other, and sequences from different songs must be far apart."\cite{ref_article21}

There is already a well-known software for music searching - Shazam\cite{ref_url6}, their method was introduced in "An Industrial Strength Audio Search Algorithm"\cite{ref_article12}. This software can search for a song, which means it listens to a song sequence and then returns the most related song to that sequence. The query sequences have to be audio recorded songs, which also works with some of the remixes and covers. Shazam\cite{ref_url6} "listens" to a song, but does not perform very well on hummed tunes.

"Hum to Search" is a feature in Google search whose methodology was introduced in a blog\cite{ref_url4} in 2020. This search feature allows you to hum your tune and returns the top likely options based on the tune, and then you can choose what matches the best for you (like a recommended system). However, as mentioned in the Introduction[\ref{introduction}], they cannot return accurate results when searching for a Vietnamese song. 

The research is take place from Zalo AI Challenge 2021\cite{ref_url1}'s first prize solution, the solution was created and published by Wano (a three-member team consisting of Mr Vo Van Phuc, Mr Nguyen Van Thieu, and Mr Lam Ba Thinh) on a Github repository named hum2song\cite{ref_url3}.

\section{Methodology}\label{methodology}
The methodology includes 3 parts: data preprocessing, training embedding model, and inference.  The preprocessing of all data for training and inference is only done once. The training embedding models phase is experimented with various state of the arts to find the most sufficient backbone for the task. The inference phase is handled by using Faiss\cite{ref_article2} module. 
% This paper ran experiments on the training, the data preprocessing and inference using the same hyperparameters and config.

\subsection{Data Preprocessing}

% \subsection{Observation about the data}
\textbf{Observation about the data}

The data consists of 3 sets. The training set has 1000 unique song sequences with their unique id within 2901 song sequences along with 2901 hummed tunes, totals of 1.3 GB. The public test used for evaluating on the public leader board has 419 song sequences and 500 hummed tunes, totals of 539MB. The private test used for the final leader board standing has 10153 song sequences and 1067 hummed tune, totals of 12.5GB. You can request the data at Zalo AI Challenge website\cite{ref_url1}.

\begin{figure}[h!]
    \centering
    \includegraphics[scale=0.75]{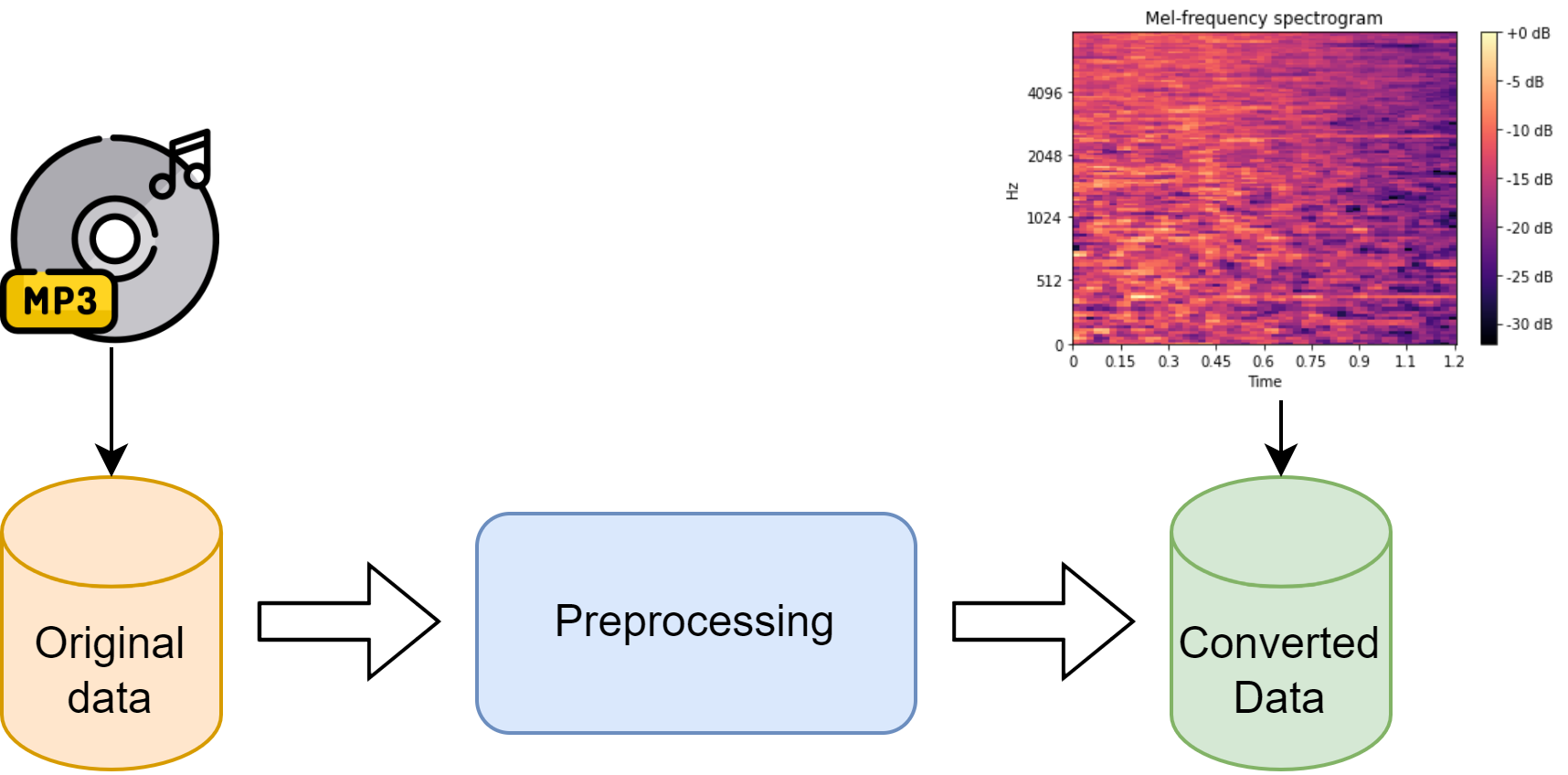}
    \caption{Preprocessing data pipeline}
    \label{fig:preprocessing_data}
\end{figure}

\noindent\textbf{Convert data from mp3 to mel-spectrogram}

As the data is original in mp3 format, in order to use it, we process all of it into mel-spectrogram and store it in numpy array format for easy training and inference. 
% The code part is adapted from Prem Seetharaman's pytorch stft\cite{ref_url2} which is a combination with librosa\cite{ref_url7} library for handling sounds in python. The data was first loaded in mp3 format using librosa\cite{ref_url7} with the sample rate of 16000, then converted to wav format.
We convert data to float32 and normalize the sound by dividing by the largest absolute value in the sound. Next, we convert to mel-spectrogram using config of filter length: 1024, hop-length: 256, win-length: 1024. Finally, we saved all preprocessed file to numpy array. The pipeline is illustrated in Figure ~\ref{fig:preprocessing_data}.

\subsection{Training}
\begin{itemize}
    % \item[$\blacksquare$] \textbf{Hardware}: Using Google Colab Pro\cite{ref_url9} with GPU Tesla P100 16GB VRAM and 16GB RAM. The training time for each backbone is recorded in Experiments\ref{experiments}. You can train with Free tier Colab with Tesla T4 or K80 but training time could be longer.

    \item[$\blacksquare$] \textbf{Hyperparameters}: All of the backbones were trained on 100 epochs, batch size of 32, the convergence of models vary depends on backbone architecture. The loss function used for this training pipeline is ArcFace, which was introduced in "ArcFace: Additive Angular Margin Loss for Deep Face Recognition" \cite{ref_article1}. The optimizer is Stochastic gradient descent with a learning rate of 1e-2, learning rate decay of 0.5 and a weight decay of 1e-1.
\end{itemize}
    
% \begin{itemize}
%     \item Epochs: 100 (Some architecture take fewer epochs to converge)
%     \item Loss function: ArcFace loss\cite{ref_article1}
%     \item Optimizer: Stochastic gradient descent
%     \item Batch size: 32
%     \item Learning rate: 1e-2
%     \item Weight decay: 1e-1
% \end{itemize}
% \subsubsection{Sample Heading (Third Level)} Only two levels of
% headings should be numbered. Lower level headings remain unnumbered;
% they are formatted as run-in headings.

% \paragraph{Sample Heading (Fourth Level)}
% The contribution should contain no more than four levels of
% headings. Table~\ref{tab1} gives a summary of all heading levels.
\subsection{Inference}
% \subsection{Inference}
% When there is a hummed tune query to the pipeline, it will be converted to mel-spectrogram and be extracted feature by embedding model. Finally, it will be used to search in our Faiss\cite{ref_article2} index (store all embedded original song sequences).

\begin{figure}[h!]
    \centering
    \includegraphics[scale=0.15]{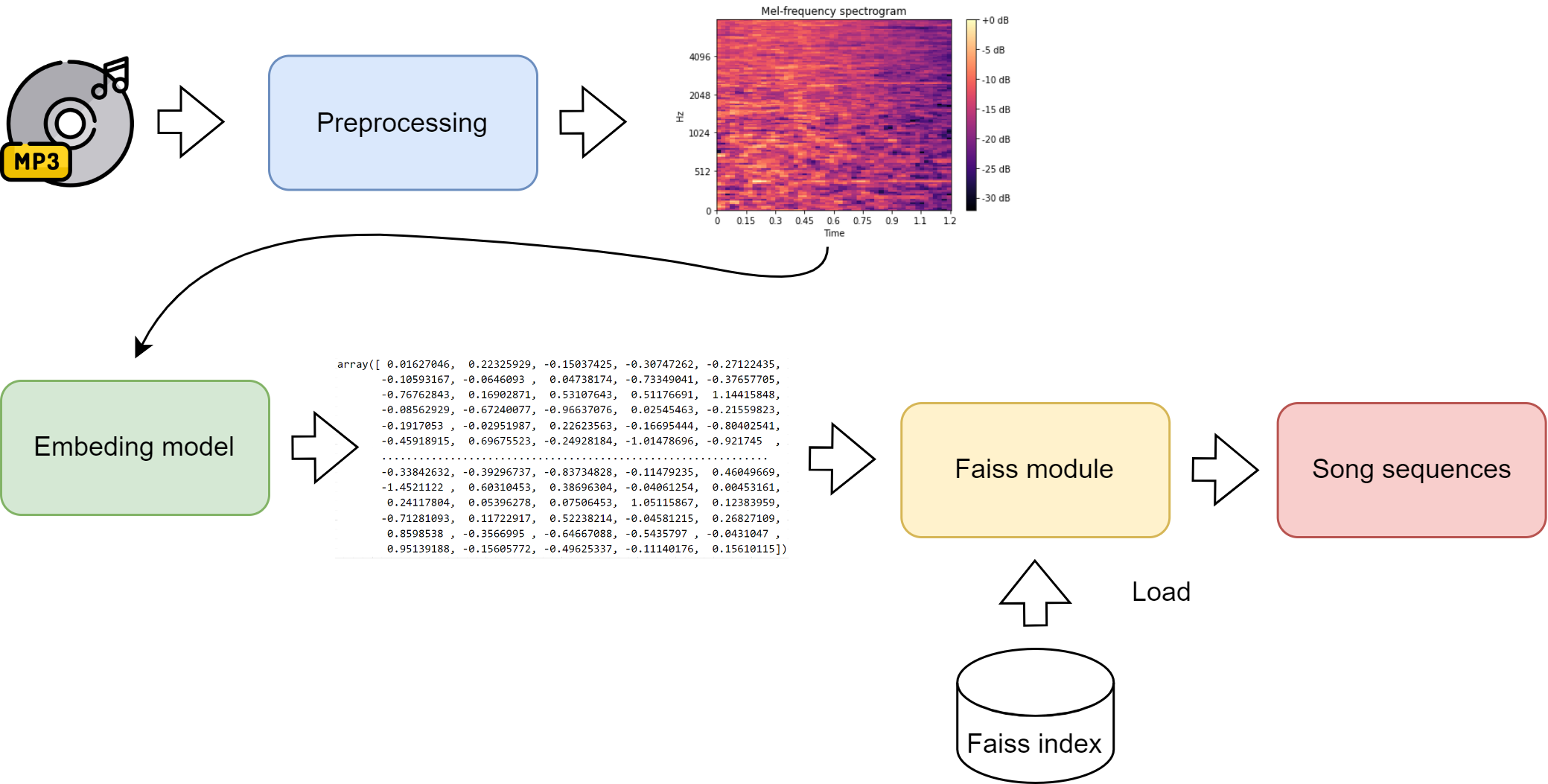}
    \caption{Inference pipeline}
    \label{fig:inference}
\end{figure}

The inference part contains 3 steps, illustrated in Figure~\ref{fig:inference}:
\begin{itemize} 
    \item Step 1: Extract all features from both song sequences.
    \item Step 2: Add song sequences' features to Faiss\cite{ref_article2} module to create a vector space of original song sequences.
    \item Step 3: Extract features from the hummed tune and use the Faiss\cite{ref_article2} module to query the closest song sequence to the hummed tune.
\end{itemize} 

The Faiss\cite{ref_article2} module using IndexFlatL2\cite{ref_article2} measures the L2 (or Euclidean) distance between all given points between our query vector, and the vectors loaded into the index. 
% To learn more about the Faiss\cite{ref_article13} module, please go to the faiss-tutorial\cite{ref_url10}, written by James Briggs.
% The competition metric, with one hummed tune we return 10 likely song sequence to that hummed tune.

% The inference time: (which averages for all implemented backbones, same Faiss\cite{ref_article2} module setup). The time for inference the public test (size 539MB) arounds 40s and private test (size 12.5Gb) around 647s.

% \begin{itemize}
%     \item public test (size 539MB) : around ~40s
%     \item private test (size 12.5GB): around ~647s
% \end{itemize}

\section{Experiments}\label{experiments}
% \subsection{Results}
After done preprocssing data, We tried on some State of the Art backbones in order to select some of the best candidate for embeddings model based on their performance on public test set and their training time.
As the Zalo company's policy, the truth label and the results were only based on the public test, which was only used for the public leader board. Please note that all of the experiments use the same configuration. The metric used for evaluating models' performance is the same as the Zalo AI Challenge, which uses MRR@10 (mean reciprocal rank).

\subsection{Compare results on different backbones and ResNet}

% \usepackage{array}

% \begin{center}
\begin{table}[h!]
\caption{Some Training results on some backbones}\label{tab1}
\centering
\setlength{\tabcolsep}{5pt}
\begin{tabular}{|c|c|c|c|c|c|}
\hline
Model & Training hours & Loss & Accuracy & MRR & Public test\\
\hline
 Resnet18 & 10.6 & 0.2736880779 & 0.8125 & 0.9982826187 & 0.9318246032 \\ 
 Alexnet & 2.017 & 1.482195854 & 0.71875 & 0.91776315067 & 0.8409246032\\  
 Vgg11 bn & 6.75 & 0.005311204121 & 0.96875 & 0.9983863532 & 0.9063388889 \\    
 Mobilenetv2 & 3.167 & 1.437324166 & 0.65625 & 0.9135797438 & 0.8921666667\\
 \hline
\end{tabular}
\end{table}
% \end{center}

Those are results on some backbones. AlexNet\cite{ref_article11} and Mobilenetv2\cite{ref_article10} have the lowest training time, but the score on the public test is not as good as ResNet\cite{ref_article4} and VGG\cite{ref_article5}. All further experiments only focus on ResNet\cite{ref_article4} and VGG\cite{ref_article5}.

% \subsection{Compare results on ResNet}
\begin{table}[h!]
\caption{Training results on Resnet}\label{tab2}
\centering
\setlength{\tabcolsep}{7pt}
\begin{tabular}{|c|c|c|c|c|c|}
\hline
Model & Training hours & Loss & Accuracy & MRR & Public test\\
\hline
 Resnet18 & 10.6 & 0.2736880779 & 0.8125 & 0.9982826187 & 0.9318246032 \\ 
 Resnet34 & 23.6 & 0.4746982753 & 0.84375 & 0.9983287229 & 0.9423634921 \\
%  Resnet50 & & & & 
 \hline
\end{tabular}
\end{table}

The experiments show that ResNet\cite{ref_article4} returned the best score on the public test, but the architecture like ResNet34\cite{ref_article4} took nearly a day for training.

\begin{table}[h!]
\caption{Training results on Modified Resnet}\label{tab3}
\centering
\setlength{\tabcolsep}{7pt}
\begin{tabular}{|c|c|c|c|c|c|}
\hline
Model & Training hours & Loss & Accuracy & MRR & Public test\\
\hline
 Resnet18 & 15.25 & 0.08737678826 & 0.90625 & 0.9983114338 & 0.9407992063 \tabularnewline
\rowcolor{lightgray} Resnet34 & 22.9 & 0.02092118189 & 0.9375 & 0.9981961734 & 0.9457190476 \\
%  Resnet50 & & & & 
 \hline
\end{tabular}
\end{table}

The highlight ResNet34 is the final model for using in our pipeline because of its performance.
The training on ResNet\cite{ref_article4} took a lot of time, our team tried to change the architecture of ResNet\cite{ref_article4} for better training time and better at creating embeddings. Instead of using the same block in pytorch model hub implementation, the experiments running on ResNet\cite{ref_article4} which use variants from the paper Additive Angular Margin Loss for Deep Face Detection\cite{ref_article1} increase the training hours but the accuracy increased ~9, 10\% and the public test score increased ~1\% at the ResNet18\cite{ref_article4}.

\subsection{Compare results on VGG}

\begin{table}[ht!]
\caption{Training results on Vgg}\label{tab4}
\centering
\setlength{\tabcolsep}{7pt}
\begin{tabular}{|c|c|c|c|c|c|}
\hline
Model & Training hours & Loss & Accuracy & MRR & Public test\\
\hline
 VGG11 & 6.75 & 0.0053112041 & 0.96875 & 0.9983863532 & 0.9063388889 \\ 
 VGG13 & 8.233 & 0.0047464803 & 0.9375 & 0.998340249 & 0.9183611111 \\
 VGG16 & 11.1 & 0.0047137937 & 0.96875 & 0.9982249885 & 0.9229555556\\
 \hline
\end{tabular}
\end{table}
% VGG variants give comparable results with ~0.9 MRR@10 with only 6.75 hours to train.
% VGG\cite{ref_article5} network has much lower training hours than ResNet\cite{ref_article4}, but public test score on MRR@10 is lower than ResNet\cite{ref_article4} about ~3,4\%.

\subsection{Evaluation}

ResNet\cite{ref_article4} variants have more training time and 
As the results shown on the tables above, based on score and training hours trade-off, if it about the mean reciprocal rank or accuracy, ResNet\cite{ref_article4} variants and other big networks which create much better feature extraction are recommended. But as the speed of training, which also provides an acceptable score, VGG\cite{ref_article5} variants are recommended.

\begin{figure}[h!]
    \centering
    \includegraphics[scale=0.25]{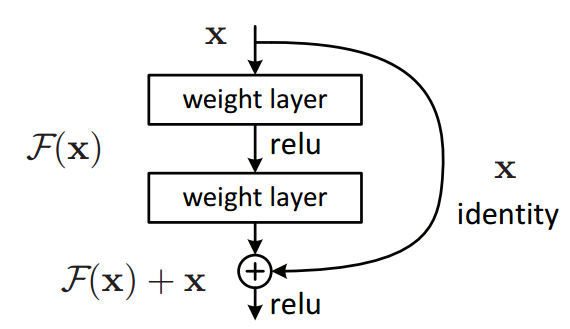}
    \caption{Residual Block — Image is taken from the original paper}
    \label{fig:ResidualBlock}
\end{figure}

VGG\cite{ref_article5} variants have gradients vanishing problem. To solve that, ResNet\cite{ref_article4} variants have "short cut connection", and instead of learning the mapping from x →F(x), the network learns the mapping from x → F(x)+G(x), showed in Figure~\ref{fig:ResidualBlock}.
% That is the reason why ResNet\cite{ref_article4} variants have more training time and better scoring than the VGG\cite{ref_article5}.

\section{Conclusion}\label{conclusion}
In this paper, we presented our pipeline for preprocessing song sequences and hummed tunes, training embedding models, and using Faiss module to optimize searching for the candidate for a hummed tune. The result on the public leader board is about ~90\%.

There are environmental and human factors in the hummed tune, such as the hummed tune contains noises or depends on the human's memory. In the preprocessing and augmentation phases, you can try to modify the hyperparameters, like increasing the sampling rate, max wav value, etc.

% There is a correlation between the size of the model to its performance in creating features from the song sequences and hummed tunes. We also want to try experiments on , ResNext, Regnet,etc. To further experiments, we suggest trying models which are one evaluated on ImageNet and on ImageNet Classification LeaderBoard.

In the searching candidate phase, Faiss module also provides ways to optimize searching. We suggest trying partitioning the index to increase the speed. When the dataset is huge and we store all features in original shapes or vectors as full (e.g Flat), Product Quantization is also supported by Faiss module.

%%===========================================================================================%%
%% If you are submitting to one of the Nature Portfolio journals, using the eJP submission   %%
%% system, please include the references within the manuscript file itself. You may do this  %%
%% by copying the reference list from your .bbl file, paste it into the main manuscript .tex %%
%% file, and delete the associated \verb+\bibliography+ commands.                            %%
%%===========================================================================================%%

% \bibliography{}% common bib file
%% if required, the content of .bbl file can be included here once bbl is generated
%%\input sn-article.bbl

%% Default %%

\end{document}